%% file: main.tex
\documentclass[sigconf]{acmart}
\usepackage{enumitem} % for itemize
\usepackage{algcompatible}
\usepackage{algorithm, algorithmicx}
\usepackage{amsmath, amsthm}
\usepackage{subfig}% for subfloat
\usepackage{multirow}% for mul row table

\AtBeginDocument{%
  \providecommand\BibTeX{{%
    \normalfont B\kern-0.5em{\scshape i\kern-0.25em b}\kern-0.8em\TeX}}}

\copyrightyear{2021}
\acmYear{2021}
\setcopyright{acmcopyright}\acmConference[KDD '21]{Proceedings of the 27th ACM SIGKDD Conference on Knowledge Discovery and Data Mining}{August 14--18, 2021}{Virtual Event, Singapore}
\acmBooktitle{Proceedings of the 27th ACM SIGKDD Conference on Knowledge Discovery and Data Mining (KDD '21), August 14--18, 2021, Virtual Event, Singapore} 
\acmPrice{15.00}
\acmDOI{10.1145/3447548.3467428}
\acmISBN{978-1-4503-8332-5/21/08}

\begin{document}
\fancyhead{}
\title[LinearModels]{Towards a Better Understanding of Linear Models \\ for Recommendation}


\author{Ruoming Jin}
\email{rjin1@kent.edu}
\affiliation{%
  \institution{Kent State University}
  \country{USA}
}

\author{Dong Li}
\email{dli12@kent.edu}
\affiliation{%
  \institution{Kent State University}
  \country{USA}
}

\author{Jing Gao}
\email{jgao@ilambda.com}
\affiliation{%
  \institution{iLambda}
  \country{USA}
}

\author{Zhi Liu}
\email{zliu@ilambda.com}
\affiliation{%
  \institution{iLambda}
  \country{USA}
}

\author{Li Chen}
\email{lchen@ilambda.com}
\affiliation{%
  \institution{iLambda}
  \country{USA}
}

\author{Yang Zhou}
\email{yangzhou@auburn.edu}
\affiliation{%
  \institution{Auburn University }
  \country{USA}
}




\input{text/abstract.tex}


\maketitle
\input{text/intro.tex}

\input{text/background.tex}

 \input{text/PCA.tex}

\input{text/Boost.tex}

 \input{text/eval.tex}
 \input{text/conclusion.tex}

 \balance
\bibliographystyle{plain}
\bibliography{bib/recommendation.bib} 
\section{Acknowledgments}
The research was partially supported by a sponsorship research agreement between Kent State University and iLambda, Inc.
\newpage
\input{text/appendix_new.tex}

\end{document}

%% file: text/abstract.tex
\begin{abstract}
Recently, linear regression models have shown to often produce rather competitive results against more sophisticated deep learning models. Meanwhile, the (weighted) matrix factorization approaches have been popular choices for recommendation in the past and widely adopted in the industry. In this work, we aim to theoretically understand the relationship between these two approaches, which are the cornerstones of model-based recommendations. 
Through the derivation and analysis of the closed-form solutions for two basic regression and matrix factorization approaches, we found these two approaches are indeed inherently related but also diverge in how they ``scale-down'' the singular values of the original user-item interaction matrix. 
We further introduce a new learning algorithm in searching (hyper)parameters for the closed-form solution and utilize it to discover the {\em nearby} models of the existing solutions.  The experimental results demonstrate that the basic models and their closed-form solutions are indeed quite competitive against the state-of-the-art models, thus, confirming the validity of studying the basic models. The effectiveness of exploring the nearby models are also experimentally validated. 
\end{abstract}

%% file: text/intro.tex
\section{Introduction}

% Recommendation systems (RS) have become a vital and integral part of the modern web, enabling the key services of both internet giants, such as Amazon, Netflix, Alibaba, etc., and millions of smaller e-commerce and online media websites. In order to better serve their customers, and to gain and maintain their competitive advantages, these companies need to continue to invest in and to improve their recommendation systems. The importance of recommendation systems is found in the core premise of the internet: connecting users with the right products, information, and/or other users. When a website offers tens of thousands to multiple billions of products, it simply goes beyond an individual customer's capability to browse and search. Thus, recommendation, or personalization, which aims to best match the preferences and/or needs of an individual customer across all the available choices, is simply indispensable. 
% Concretely, a recommendation system computes a list of ``personalized'' items pertaining to each individual user, based on her or his profile and past behaviors, such as view/click histories, etc. The typical strategy of recommendation ranks all available items, according to certain relevance scores, based on collaborative filtering and contents, and selects the top-$k$ items. 

Over the last 25 years, we have witnessed a blossom of recommendation algorithms being proposed and developed\cite{charubook,zhang2019deep}. 
Though the number of (top-n) recommendation algorithms is fairly large, the approaches can be largely classified as neighborhood approaches (including the regression-based approaches), matrix factorization (or latent factors) approaches,  more recent deep learning based approaches, their probabilistic variants, and others\cite{charubook,zhang2019deep}. 
However, there have been some interesting debates on the results being reported by recent deep learning-based approaches:  the experimental results show most of these methods seem to achieve sub-par results when compared with their simple/nonlinear counterparts~\cite{RecSys19Evaluation}. This issue also relates to selecting and tuning baselines ~\cite{Steffen@19DBLP} as well as the choice of evaluation metrics ~\cite{WalidRendle20}, among others. But recent studies also seem to confirm the state-of-the-art linear models, such as SLIM~\cite{slim01} and EASE~\cite{Steck_2019} do obtain rather remarkable results when compared to the more sophisticated counterparts~\cite{DacremaBJ21}.  

% SLIM and EASE can be considered the linear version of auto-encoders~\cite{DBLP:conf/nips/Steck20}: it utilizes a rather simple objective to search an item-to-item similarity matrix $W$, which minimizes $||X-XW||_F^2$ ($||\cdot||_F$ is the Frobenius norm) with different regularizations on $W$ and enforcing a zero diagonal for better generalization (avoiding the simple identity matrix $I$ solution). 
% In the meantime, the matrix factorization methods, such as ALS~\cite{hu2008collaborative} and SVD based approaches~\cite{mfsurvey} have been heavily favorite and widely adopted in industry for recommendation. Here the (simplified) goal is to identify matrices $U$ and $V$, such that $||X-UV||_F^2$ can be recovered together with regularizations on $U$ and $V$. The refined recommendation algorithms~\cite{hu2008collaborative} adopt the weighted version by superimposing the weight matrix $||B \cdot (X-UV)||^2_F$ ($\odot$ as the element-wise matrix multiplication). The significance of the matrix factorization methods also imply the exists of latent factors, and can be considered as embedding users and items into a higher dimensional space.  
% Interestingly, there have been a list of {\em low-rank} regression approaches~\cite{DBLP:conf/pakdd/ChristakopoulouK14} which aims to factorize $W$ as $AB$, such that $||X-XAB||_F^2$ is minimized. 

Intuitively, SLIM and EASE search an item-to-item similarity matrix $W$ so that the user-item interaction (denoted as matrix $X$ with rows and columns corresponding to users and items, respectively) can be recovered by the matrix product: $XW$. In fact, they can be considered as simplified linear auto-encoders~\cite{DBLP:conf/nips/Steck20}, where $W$ serves as both encoder and decoder, denoted as $L_W$, such that $L_W(x_u)=x_u W$ can be used to recover $x_u$ ($x_u$ is the $u$-th row vector of user-item interaction matrix $X$).  
In the meantime, the matrix factorization methods, such as ALS~\cite{hu2008collaborative} and SVD-based approaches~\cite{mfsurvey} have been heavily favored and widely adopted in industry for recommendation. They aim to discover user and item embedding matrices $P$ and $Q$, where $p_u$, $q_i$ represents the latent factors of user $u$ and item $i$, respectively, such that the user item interaction $x_{ui}$ can be approximated by $q_i^T p_u$.  
Furthermore, there have been a list of {\em low-rank} regression approaches~\cite{DBLP:conf/pakdd/ChristakopoulouK14,LFlow16,DBLP:conf/nips/Steck20} which aim to factorize the similarity matrix $W$ as $AB^T$, such that $XAB^T$ can be used to recover $X$. Here, $XA$ and $B$ also introduces the user and item  matrices, respectively (similar to matrix factorization).

Thus, on the surface, we can see the (reduced-rank) regression is like a special case of matrix factorization~\cite{fism13,LRec16}, and it also has seemingly smaller number of parameters (the size of similarity matrix $W$ is typically much smaller than the user and item latent factor matrix as the number of users tend to be much larger than items). Therefore, the expectation is the regression models are more restricted, and thus less flexible (and expressive) than the matrix factorization approaches. However, the recent results seem to indicate the regression approaches tend to perform better than the matrix factorization approaches in terms of commonly used evaluation criteria, such as Recall and nDCG~\cite{Steck_2019,DBLP:conf/nips/Steck20,DacremaBJ21}.

Is there any underlying factor/reason for the regression approaches to perform better than the matrix factorization approaches?  If and how these two different factorization (low-rank regression vs matrix factorization) relate to one another? To seek the connection and be able to compare/analyze their inherent advantage/weakness, can we unify them under the same framework? 
%This can be the key to help better understand their working mechanisms and their inherent weakness or advantages?  
As most of the deep learning, probabilistic and non-linear approaches all have the core representation  from either factorization~\cite{he2017neural} or auto-encoders~\cite{liang2018variational}, the answer to these questions will not only help understand two of the (arguably) most important recommendation methodologies: neighborhood vs matrix factorization, but also help design more sophisticated deep learning based approaches.  To our surprise, the theoretical analyses of these approaches are still lacking and the aforementioned questions remain unanswered. 

In this paper, by analyzing two basic (low-rank) regression and matrix factorization models, we are able to derive and compare their closed-form solutions in a unified framework. We reveal that both methods essentially ``scale-down'' their singular values of user-item interaction matrix using slightly different mechanisms. The (low-rank) regression mechanism allows the use of more {\em principal components} (latent dimensions) of the user-item matrix $X$ than the matrix factorization approaches. Thus, this potentially provides an inherent advantage to the former methods over the latter. Another surprising discovery is that although the matrix factorization seems to have more model parameters with both user and item latent factor matrices, its optimal solution for the simplified problem suggests that it is actually only dependent on the item matrix. Thus, it is actually more restricted and less flexible than the regression based approaches.  This again indicates the potential disadvantage of the matrix factorization approaches. 
%that the matrix factorization is potentially no more (or even less) flexible than the regression based approaches. 

To help further understand how the singular values of user-item interaction matrix can be adjusted (at individual level), we introduce a novel learning algorithm which can search through high dimensional continuous (hyper)parameter space. This learning algorithm also enables us to perform the post-model fitting exploration ~\cite{guan2018post} for existing linear recommendation models. Our approach is to augment (existing) linear models with additional parameters to help further improve the model accuracy. The resulting models remaining as linear models, which can be considered as {\em nearby} models with respect to the existing models. 
This approach indeed shares the similar spirit of the our recently proposed ``next-door analysis'' from statistical learning~\cite{hastie@statisticallearning} though our approaches and targets are quite different. 
To the best of our knowledge, this is the first work to study post-model fitting exploration for recommendation models. Such study can not only help better evaluate the optimality of the learned models, but also (potentially) produce additional boost for the learned models. 

As pointed out by ~\cite{DacremaBJ21}, a major problem in existing recommendation research is that the authors tend to focus on developing new methods or variants of recommendation algorithms, and then validate based on ``hyper-focus on abstract metrics" with often weak or not-fully-tuned baselines to ``prove'' the progress. Though using better baselines, datasets, and evaluation metrics can help address of part of the problem, a better understanding of how, why, and where the improvement over existing are being made is equally important. We hope the theoretical analysis, the new learning tool for (hyper)parameter search, and the post-model analysis on linear recommendation models can be  part of the remedy for the aforementioned problem.

To sum, in this paper, we made the following contribution: 
\begin{itemize}[leftmargin=*,noitemsep,nolistsep]
\item (Section~\ref{pca}) We theoretically investigate the relationship between the reduced-rank regression (neighborhood) approaches and the popular matrix factorization approaches using the closed-form solutions, and reveal how they connect with one another naturally (through the lens of well-known principal component analysis and SVD). We also discover some potential factors which may provide a benefit for the regression based methods.   
\item (Section~\ref{learning}) We introduce a new learning algorithm to help search the high-dimension (hyper)parameter space for the closed-form solution from Section~\ref{pca}. We further apply the learning algorithm to perform post-model exploration analysis on the existing linear models by augmenting them with additional parameters (as nearby models).   
\item (Section~\ref{experiments}) We experimentally validate the closed-form solution for the basic regression and matrix factorization models, and show their (surprising) effectiveness and accuracy comparing against the state-of-the-art linear models; we also experimentaly validate the effectiveness of the learning algorithms for the closed-form solutions and identifying nearby models. We show nearby models can indeed boost the existing models in certain datasets. 
 \end{itemize}

%% file: text/background.tex
\vspace*{-2.0ex}
\section{Background}
\label{problem}
Let the training dataset consists of $m$ users and $n=|I|$ items, where $I$ is the entire set of items. In this paper, we will focus on the implicit setting for recommendation. Compared with the explicit settings, the implicit setting has more applications in ecommerce, content recommendation, advertisement, among others. It has also been the main subjects for recent top-$n$ recommendation~\cite{Steck_2019,hu2008collaborative,slim01, RecSys19Evaluation,zhang2019deep}.
%since the Netflix competition which has explicit ratings. 
Here, the user-item interaction matrix $X$ can be considered as a binary matrix, where $x_{ui}=1$ represents there is an interaction between user $u$ and item $i$. If there is no interaction between $u$ and $i$, then $x_{ui}=0$. 
Let $X_u^+=\{j: x_{uj}>0\}$ denote the item set that user $u$ has interacted with, and $X_u^-=I-X_u^+$ to be the item set that $u$ has not interacted with. 

\subsection{Regression Models for Neighborhood-based Recommendation}
It is well-known that there are user-based and item-based neighborhood based collaborative filtering, and the item-based approach has shown to be more effective and accurate compared with user-based approaches~\cite{DeshpandeK@itemKNN}. Thus, most of the linear models are item-based collaborative filtering (ICF). 

Intuitively, the model-based ICF aims to predict $x_{ui}$ (user $u$'s likelihood of interaction with and/or preference of item $i$) based on user $u$'s past interaction with other items $X_u^+$:
\begin{equation}
\hat{x}_{ui}=\sum_{j \in X_u^+} s_{ji} x_{uj}, 
\end{equation}
where $s_{ji}$ denotes the similarity between item $j$ and $i$. 

The initial neighborhood approach uses the statistical measures, such as Pearson correlation and cosine similarity ~\cite{charubook} between the two columns $X_{*i}$ and $X_{*j}$ from items $i$ and $j$. The more recent approaches have been aiming to use a regression approach to directly learn the weight matrix $W$ (which can be considered as the inferred similarity matrix)  so that $||X-XW||_F^2$ ($||\cdot||_F$ denotes the Frobenius norm) is minimized. Clearly, in this formulation, the default solution $W=I$ should be avoided for generalization purpose, and the difference of different approaches lie in the constraints and regularization putting on $W$. Recent studies have shown these approaches achieve comparable or even better performance compared with the state-of-the-art deep learning based approaches ~\cite{DBLP:conf/nips/Steck20,DacremaBJ21}. 

\noindent{\bf SLIM:}
SLIM~\cite{slim01} is one of the first regression-based approach to infer the weight matrix $W$. It considers $W$ to be nonnegative, and regularizing it with $L_1$ and $L_2$ norm (thus {\em ElasticNet})~\cite{elasticnet05}. In addition, $S$ require to be zero diagonal:
\begin{equation}
\begin{split}
    W & =\arg \min_{W}   \frac{1}{2}||X-XW||^2_F+ \lambda_1||W||_1 + \lambda_2 ||W||_F^2 \\ 
  & s.t.\ \  W\geq 0, diag(W)=0, \\
\end{split}
\end{equation} 
where $||\cdot||_1$ denotes the $L_1$ matrix norm, and $diag(\cdot)$ denotes the diagonal (vector) of the corresponding matrix. 
Since no closed-form solution for $W$, the solver of ElasticNet is used to optimize $W$, and $\lambda_1$ and $\lambda_2$ are the correspondingly regularization hyperparameters. 

There are a quite few variants of SLIM being proposed, including HOLISM~\cite{hoslim14} which extends SLIM to capture higher-order relationship, and LRec~\cite{LRec16} which considers a non-linear logistic loss (instead of squared loss) with no zero diagonal, no negative and $L1$ constraints, among others.  

\noindent{\bf EASE:}
EASE~\cite{Steck_2019} is a recent regression-based model which has shown to improve over SLIM with both speed and accuracy, and quite competitive again the state-of-the-art deep learning models. It simplifies the constraint and regularization enforced by SLIM by removing non-negative and $L1$ constraints: 
\begin{equation}
\begin{split}
W & =\arg \min_{W}   \frac{1}{2}||X-XW||^2_F+ \lambda ||W||_F^2 \\ 
& s.t. \quad diag(W)=0\\
\end{split}
\end{equation} 
Empirical study ~\cite{Steck_2019} basically confirms that the non-negative constraint and $L_1$ norm on matrix $W$ may not be essential (or have negative impact) on the performance. 
Particularly, EASE has a closed-form solution~\cite{Steck_2019}. 
% $\overhat{W}_{EASE} =$
% \begin{equation}
%  P ^{-1} (P - diagMat ({\bf \vec{1}} \oslash diag(P)) = I- P diagMat ({\bf \vec{1}} \oslash diag(P)),
% \end{equation}
% where $P=(X^T X+ \lambda I)^{-1}$, and $\oslash$ denotes the element-wise division, $diagM(\cdot)$ denotes the diagonal matrix. 

\noindent{\bf DLAE and EDLAE:}
The latest extension of EASE, the DLAE (Denoising linear autoencoder) ~\cite{DBLP:conf/nips/Steck20} utilizes a drop-out induced the $L_2$ norm to replace the standard $L_2$ norm without zero diagonal constraints: 
\begin{equation} \label{eq:DLAE}
W =\arg\min_{W}\frac{1}{2}||X-XW||^2_F+ ||\Lambda^{1/2}W||_F^2 
\end{equation} 
where $\Lambda=\frac{p}{1-p} diagM(diag(X^T X))$ ($diagM(\cdot)$ denotes the diagonal matrix) and $p$ is the dropout probability. 

% The optimal solution has the closed form $\overhat{W}_{DLAE}=$ 
% \begin{equation}
% PX^TX = (X^X+\Lambda)^{-1}(X^TX+\Lambda-\Lambda)=I-(X^TX+\Lambda)^{-1}\Lambda.
% \end{equation}

Another variant EDLAE would explicitly enforce the zero diagonal constraints: 
\begin{equation}
\begin{split}
    W & =\arg \min_{W}   \frac{1}{2}||X-XW||^2_F+ ||\Lambda^{1/2}W||_F^2 \\ 
  & s.t. \ \  diag(W)=0, \\
  \end{split}
\end{equation}
Both DLAE and EDLAE have closed-form solutions~\cite{DBLP:conf/nips/Steck20}. 
%It optimal solution $\overhat{W}_{DLAE}$ has the same format as EASE with a different $P=(X^X+\Lambda)^{-1}$. 

\subsubsection{Low-Rank Regression}
There have been a number of interesting studies ~\cite{fism13,LRec16,DBLP:conf/nips/Steck20} on using low-rank regression to factorize the weight/similarity matrix $W$. The latest work ~\cite{DBLP:conf/nips/Steck20} shows a variety of low-rank regression constraints which have been (or can be) used for this purpose: 
\begin{equation}
\begin{split}
||X-XAB^T||_F^2 & +\lambda (||A||_F^2+||B^T||_F^2) \\
    ||X-XAB^T||_F^2 & +\lambda ||AB^T||_F^2 \\
   ||X-XAB^T||_F^2 & + ||(\Lambda+\lambda I)AB^T||_F^2 
\end{split}
\end{equation}
where $A_{n\times k}$,  $B_{n\times k}$, and thus $rank(AB) \leq k$. 
The reduced-rank EDLAE ~\cite{DBLP:conf/nips/Steck20} further enforces zero diagonal constraints for generalization purpose.  

We note that interestingly, the reduced-rank regression solution naturally introduces a $k$-dimensional vector embedding for each user from $XA$ ($m \times k$) , and a $k$-dimensional vector embedding for each item via $B$. This immediately leads to an important question: how such embedding differs from the traditional matrix factorization (MF) approaches which aim to  explicitly decompose $X$ into two latent factor matrices (for users and items). Note that in the past, the MF methods are more popular and widely used in industry, but the recent researches ~\cite{DacremaBJ21} seem to suggest an edge based on the regression-based (or linear autoencoder) approaches over the MF approaches. Before we formalize our question, let us take a quick review of  MF approaches. 

\subsection{Matrix Factorization Approaches}
Matrix factorization has been been widely studied for recommendation and is likely the most popular recommendation methods (particularly showing great success in the Netflix competition~\cite{mfsurvey}). 
The basic formula to estimate the rating is 
\begin{equation}
    \hat{x}_{ui}= p_u \cdot q_i = q_i^T p_u, 
\end{equation}
where $p_u$ and $q_i$ are the corresponding $k$-dimensional latent vectors of user $u$ and item $i$, respectively. 
Below, we review several well-known matrix factorization approaches for implicit settings; they differ on how to treat known vs missing interactions and regularization terms, among others.

\noindent{\bf ALS:}
The implicit Alternating Least Square (ALS) method~\cite{hu2008collaborative} is basically a weighted matrix factorization (WRMF): 
\begin{equation}\label{eq:WMFALS}
     \arg \min_{P,Q} ||C \odot (X- PQ^T)||_F^2 + \lambda (||P||_F^2 + ||Q||_F^2), 
\end{equation}
where $P_{m \times k}$ records all the $k$-dimensional latent vectors for users and $Q_{n \times k}$ records all the item latent vectors, and $\lambda$ regularize  the squared Frobenius norm. $C$ is the weight matrix (for binary data, the known score in $X$ typically has $\alpha$ weight and the missing value has $1$), and $\odot$ is the element-wise product. For the general weight matrix, there is no closed-form solution; the authors thus propose using alternating least square solver to optimize the objective function. 

\noindent{\bf PureSVD:}
The Matrix factorization approach is not only closely related to SVD (singular value decomposition), it is actually inspired by it~\cite{charubook}. In the PureSVD approach, the interaction matrix $X$ is factorized using SVD (due to Eckart-Young theorem)~\cite{CremonesiKT@10}: 
%\vspace{-1.ex}
\begin{equation}
    \arg \min_{U_k,\Sigma_k,V_k} ||X-U_k \Sigma_k V_k^T||_F^2,
\end{equation}
where $U_k$ is a $m \times k$ orthonormal matrix, $V_k$ is a $n \times k$ orthonormal matrix, and $\Sigma_k$ is a $k \times k$ diagonal matrix containing the first $k$ singular values. 
Thus the user factor matrix can be defined as $P=U_k \Sigma_k$ and the item factor matrix is $Q=V_k$. 

\noindent{\bf SVD++:}
SVD++ ~\cite{Koren08} is another influential matrix factorization approach which also integrate the neighborhood factor. It nicely combines the formulas of  factorization and neighborhood approaches with generalization. It targets only positive user-item ratings and typically works on explicit rating prediction. 

\subsection{The Problem}
As we mentioned earlier, a few recent studies ~\cite{DBLP:conf/nips/Steck20,DacremaBJ21} seem to indicate the regression based approach (or linear autoencoder approach) seem to have better performance than the popular matrix factorization approach, such as ALS~\cite{HuSWY18}. However, if we look at the reduced rank regression approach, we observe its solution can be considered a special case of matrix factorization. Another interesting question is on the regularization hyperparameter, $\lambda$: ALS typically use a much smaller regularization penalty compared with the one used in the regression based approach, such as EASE and low-rank version. The latter's  $\lambda$ value is typically very large, in the range of thousands and even tens of thousands or ~\cite{Steck_2019,DBLP:conf/nips/Steck20}. Note that both aim to regularize the squared Frobenius matrix norm. What results in such discrepancy?  

Another interesting problem is about model complexity of these two approaches. The regression-based (linear auto-encoder) approach uses the similarity matrix $W$ (which has $n\times n$ parameters), and when using low-rank regression, its parameters will be further reduced to $O(n \times k)$ where $k$ is the reduced rank. The MF has both user and item matrices, and thus has $O((m+n)k)$ parameters. This seems to indicate the MF approach should be more flexible than the regression approaches as it tends to have much more parameters (due to number of users is typically much larger than the number of items). But is this the case?   

% Furthermore, both methods have non-linear, probabilistic and deep learning variants which at the core are either utilizing the autoencoder framework (each user's past behavior as input to generate user embedding), such as DMF and Multi-VAE~\cite{} or explicit consider user and item embedding, such as NCF~\cite{}. 

In this study, our focus is not to experimentally compare these two types of recommendation approaches, but instead to have a better theoretical understanding their differences as well their connections. Thus, we hope to  understand why and how if any approach maybe more advantageous than the other and along this, we will also investigate why the effective range of their regularization hyper-parameters are so different.
We will also investigate how to learn high dimensional (hyper)parameters and apply it to help perform post-model exploration to learn "nearby" models.

%% file: text/PCA.tex
\vspace*{-1.0ex}
\section{Theoretical Analysis} 
\label{pca}
In this section, we will theoretically investigate the regression and matrix factorization models, and explore their underlying relationships, model complexity, and explain their discrepancy on regularization parameters. 

\vspace*{-1.0ex}
\subsection{Low-Rank Regression (LRR) Models}
To facilitate our discussion, we consider the following basic low-rank regression models: 

\begin{equation}
\label{eq:lowrank}
    W=\arg \min_{rank(W)\leq k} ||X -XW||_F^2+ 
    ||\mathbf{\Gamma} W||_F^2,
\end{equation}
where  $\mathbf{\Gamma}$ Matrix regularizes the squared Frobenius norm of $W$. (This can be considered as the generalized ridge regression, or multivariant Tikhonov regularization)~\cite{vanwieringen2020lecture}.
%$\mathbf{\Gamma}^T \mathbf{\Gamma}$ is a positive definite matrix and symmetric matrix, and 
For the basic case, $\mathbf{\Gamma}^T \mathbf{\Gamma} =\lambda I$, and $\mathbf{\Gamma}^T \mathbf{\Gamma} =\Lambda=\frac{p}{1-p} diagM(diag(X^T X))$ for DLAE (eq ~\ref{eq:DLAE}) ~\cite{DBLP:conf/nips/Steck20}.
Note that this regularization does not include the zero diagonal requirement for $W$. As we will show in Section~\ref{experiments}, enforcing it only provides minor improvement and thus the basic model can well capture  the essence of (low-rank) regression based recommendation. 

%A popular of the item reweight $d_i$ is the inverse of the squared root of item popularity~\cite{}.  In the following analysis, we will consider the generic setting of hyper-parameters $d$ and $\mathbf{\Gamma}$.

To help derive the closed-form solution for above problem, let us represent it as a standard regression problem. 
\begin{equation*}
\overline{Y}= \begin{bmatrix} X  \\ 0    \end{bmatrix} \ \ \ \ \overline{X}=\begin{bmatrix} X  \\ \boldsymbol{\Gamma}   \end{bmatrix}
\end{equation*}

Given this, the original problem can be rewritten as:
\begin{equation*} 
\begin{split}
    \min_{rank(W)\leq k}& ||\overline{Y}-\overline{X}W||_F^2 = \\
    \min_{rank(W) \leq k}& ||\overline{Y}-\overline{X} W^*||_F^2 +
     ||\overline{X}W^*-\overline{X}W||_F^2, 
\end{split}
\end{equation*}
where $W^*=\arg\min ||\overline{Y}-\overline{X} W^*||_F^2$. Basically, the initial loss $||\overline{Y}-\overline{X}W||_F^2$ is decomposed into two parts: $||\overline{Y}-\overline{X} W^*||_F^2$ (no rank constraint), and $||\overline{X}W^*-\overline{X} W||_F^2$. 
Note this holds since the vector ( $\overline{Y}-\overline{X}W^*$) is orthogonal to $\overline{X}W^*-\overline{X}W=\overline{X}(W^*-W)$ (The optimality of Ordinary Least-Square estimation~\cite{vanwieringen2020lecture}).  

Now, the original problem can be broken into two subproblems: 

\noindent{\bf (Subproblem 1:) item-weighted Tikhonov regularization:}

\begin{equation*}
\begin{split}
       W^*&=\arg\min_W ||\overline{Y}-\overline{X}W^*||_F^2=\arg \min_W ||X -XW||_F^2+||\mathbf{\Gamma} W||_F^2 \\
       &= (\overline{X}^T \overline{X})^{-1}\overline{X}^T \overline{Y}  \\
       &= (X^TX+ \Gamma^T \Gamma)^{-1} X^T X
\end{split}
\end{equation*}

\noindent{\bf (Subproblem 2:) low-rank matrix approximation: }
\begin{equation*}
\begin{split}
 \hat{W} & =\arg\min_{rank(W) \leq k} ||\overline{X}W^*-\overline{X}W||_F^2  \\
   &= \arg \min_{rank(W) \leq k} ||X W^* - XW||_F^2+ 
    ||\mathbf{\Gamma} (W^*- W)||_F^2
\end{split}
\end{equation*}
Let $\overline{Y}^*=\overline{X}W^*$, and based on the well-known {\em Eckart-Young} theorem~\cite{eckart1936approximation}, we have the best rank $k$ approximation of  $\overline{Y}^*$ in Frobenius norm is best represented by SVD. 
Let $\overline{Y}^*=P \Sigma Q^T$ ($P,Q$ are orthogonal matrices and $\Sigma$ is the singular value diagonal matrix, and then the best rank $k$ approximation of $\overline{Y}^*$, denoted as $\overline{Y}^*(k)$ is 
\begin{equation}
    \overline{Y}^*(k) =P_k \Sigma_k Q^T_k,
\end{equation} 
where $M_k$ takes the first $k$ rows of matrix $M$. 
We also have the following equation: 
\begin{equation*}
    P \Sigma Q^T (Q_k Q_k^T) = P_k \Sigma_k Q^T_k
\end{equation*}
Given this, we notice that 
\begin{equation*} 
\begin{split}
    \overline{Y}^*(k) &  = P_k \Sigma_k Q^T_k  =P \Sigma Q^T (Q_k Q_k^T) \\
    & = \overline{X} W^* (Q_k Q_k^T)
      =  \overline{X} W
\end{split}
\end{equation*}
Thus, we have
\begin{equation*}
\widehat{W}=W^* (Q_k Q_k^T)=(X^TX+ \Gamma^T \Gamma)^{-1} X^T X (Q_k Q_k^T), 
\end{equation*}
and the complete estimator for $XD$ (interaction/rating inference) is written as: 
\begin{equation} 
\boxed{
\widehat{W}= (X^TX+ \Gamma^T \Gamma)^{-1} X^T X (Q_k Q_k^T)
}
\end{equation}

Next, let us further simplify it using SVD which can better reveal its ``geometric'' insight. 

\subsubsection{Eigen Simplification}
First, let the SVD of $X$ as 
\begin{equation*}
X=U \Sigma V 
\end{equation*}
When $\Gamma=\Lambda^{1/2} V^T $ where $\Lambda$ is a diagonal matrix, we can observe: 
\begin{proposition}
\begin{equation*}
Q_k=V_k
\end{equation*}
\end{proposition}

%\begin{proof}
\begin{equation*}
\overline{Y}^*=\overline{X} W^*= \overline{X} V(\Sigma^{2}+\Lambda)^{-1}\Sigma^2 V^T 
\end{equation*}
Then, from 
\begin{equation*}
\begin{split}
(\overline{Y}^*)^T \overline{Y}^* = V \Sigma^{-2} (\Sigma^{2}+\Lambda) V^T \overline{X}^T \overline{X} V(\Sigma^{2}+\Lambda)^{-1}\Sigma^2 V^T  \\ 
=V (\Sigma^2 + \Lambda) V^T 
\end{split}
\end{equation*}
%\end{proof}

Then we have the following: 
\begin{equation*}
\begin{split}
\widehat{W} &= (X^TX+ \Gamma^T \Gamma)^{-1} X^T X (Q_k Q_k^T) \\
& = V (\Sigma^2+\Lambda)^{-1}  \Sigma^2 V^T (V_k V_k^T) \\
& = V diag(\frac{\sigma^2_1}{\sigma^2_1+\lambda_1},\dots,\frac{\sigma^2_k}{\sigma^2_n+\lambda_k}) V^T (V_k V_k^T) 
\end{split}
\end{equation*}

Thus, we have the following closed-form solution:  
\begin{equation}
\label{eq:regressionlambda}
\begin{split}
\boxed{
\widehat{W} = V_k  diag(\frac{\sigma^2_1}{\sigma^2_1+\lambda_1},\dots,\frac{\sigma^2_k}{\sigma^2_k+\lambda_k})  V_k^T
}
\end{split}
\end{equation}

Now, if $\lambda_i=\lambda$, we have: 
\begin{equation}\label{eq:regressionpca}
\boxed{
\widehat{W} = V_k diag(\frac{\sigma^2_1}{\sigma^2_1+\lambda}, \cdots, \frac{\sigma^2_k}{\sigma^2_k+\lambda}) V_k^T 
}
\end{equation}
Note that this special case $\Gamma^T\Gamma=\lambda I$ has indeed been used in ~\cite{LFlow16} for implicit recommendation. However, the authors do not realize that it actually has a closed-form solution. 

We also note that using the matrix factorization perspective, we obtain the user ($P$) and item ($Q$) matrices as: 

\begin{small}
\begin{equation}
\label{eq:regressionmf}
\boxed{
\begin{split}
P=& X V_k diag(\frac{\sigma^2_1}{\sigma^2_1+\lambda},\dots,\frac{\sigma^2_k}{\sigma^2_k+\lambda})  = U_k diag(\frac{\sigma_1}{1+\lambda/\sigma^2_1},\dots,\frac{\sigma_k}{1+\lambda/\sigma^2_k}) \\
Q=& V_k 
\end{split}
}
\end{equation}
\end{small}

\subsection{Matrix Factorization}
To study the relationship between the regression approach and matrix factorization, we consider the basic regularized SVD ~\cite{zheng2018regularized}: 

\begin{equation}
\label{eq:matrixfactorization}
     \arg \min_{P,Q} || X- PQ^T||_F^2 + \lambda^\prime (||P||_F^2 + ||Q||_F^2), 
\end{equation}
The solution for this type problem is typically based on Alternating Least Square, but authors in ~\cite{zheng2018regularized} have found  a closed-form solution. 

Let $X=U \Sigma V^T$, and then let 
\begin{small}
\begin{equation}
\label{eq:mf}
\boxed{
\begin{split}
    P&=U_k diag(\sigma_1-\lambda^\prime, \cdots, \sigma_k-\lambda^\prime)   \\
    &=U_k diag(\sigma_1 (1 -\lambda^\prime/\sigma_1), \cdots, \sigma_k(1-\lambda^\prime/\sigma_k)) \\
    Q&=V_k 
\end{split}
}
\end{equation}
\end{small}

Before we further analyze the relationship between them, we ask the following interesting question: Can matrix factorization be represented as a linear encoder? In other words, we seek if there is an $W$ such that $XW=PQ$ (defined by matrix factorization). Let the Moore-Penrose inverse $X^+=V \Sigma^{-1} U^T$, then we have $\widehat{W} =X^+PQ$, 
\begin{equation}
\label{eq:mfpca}
\boxed{
\begin{split}
    \widehat{W}=V_k diag(1-\lambda^\prime/\sigma_1, 
               \cdots, 1-\lambda^\prime/\sigma_k) V_k^T 
\end{split}
}
\end{equation}

\subsection{Model Comparison and Analysis}
When $\lambda=\lambda^\prime=0$ (no regularization), then both approaches (using the matrix factorization) correspond to the standard SVD decomposition, where $P=U_k \Sigma_k$ and $Q=V_k^T$. 
Further, both approaches will  also have  $\widehat{W}=V_k V_k^T$. Then, let us consider $X\widehat{W}=X V_k V_k^T$, which is indeed our standard principal component analysis (PCA), where $V_k$ serves as a linear map which transforms each row vector $x_i^T$ to the new coordinate under the principal component basis. Clearly, when $\lambda=\lambda^\prime \neq 0$, both models start to diverge and behave differently, and results in their difference in terms of regularization penalties, model complexities and eventually model accuracy. 

\begin{figure}%
    \centering
    \subfloat[\centering scaling-down singular values \label{fig:ml-20m-sigma} ]{{\includegraphics[width=.615\linewidth]{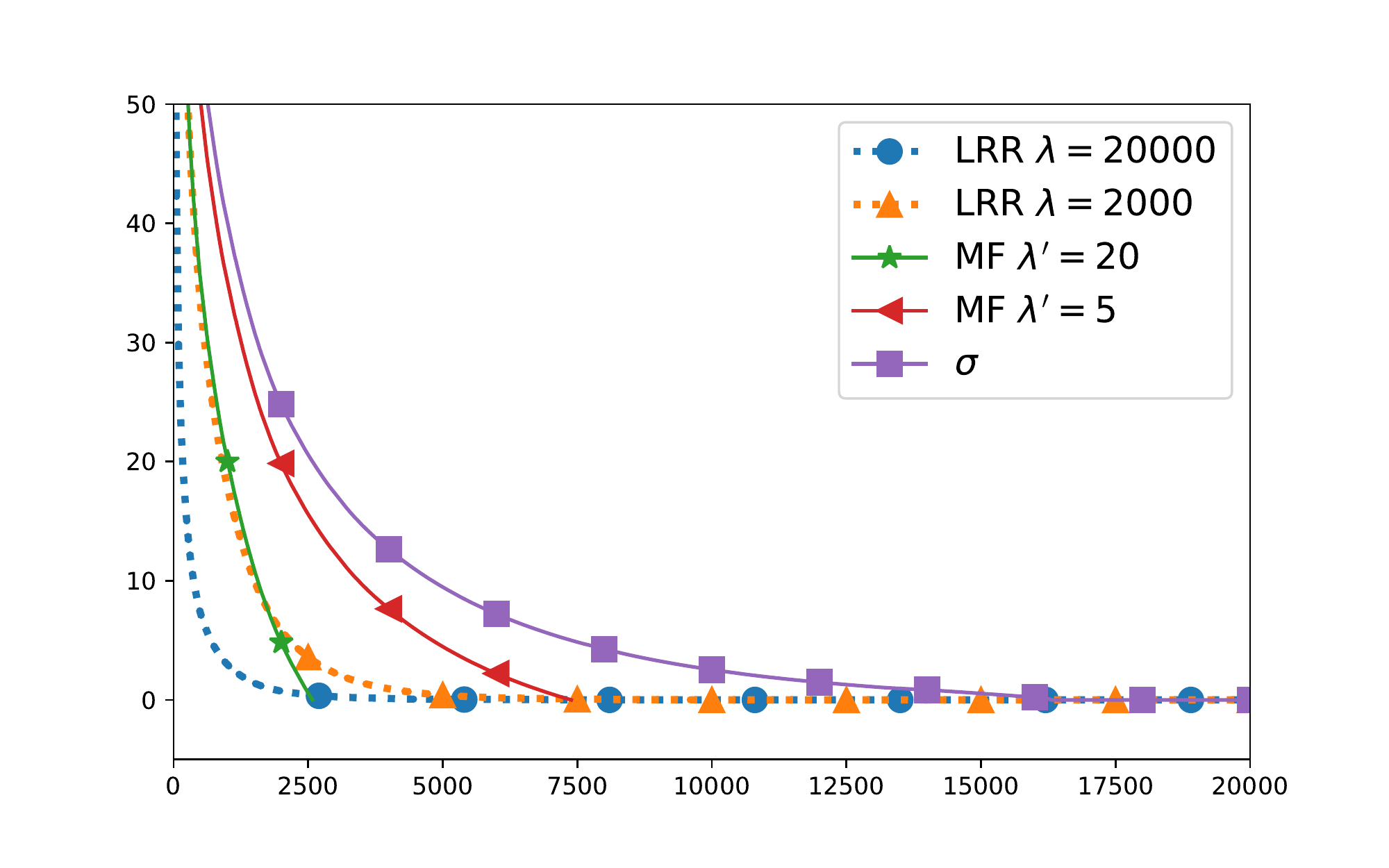} }}%
    \subfloat[\centering compression ratios \label{fig:ml-20m-pca}]{{\includegraphics[width=.385\linewidth]{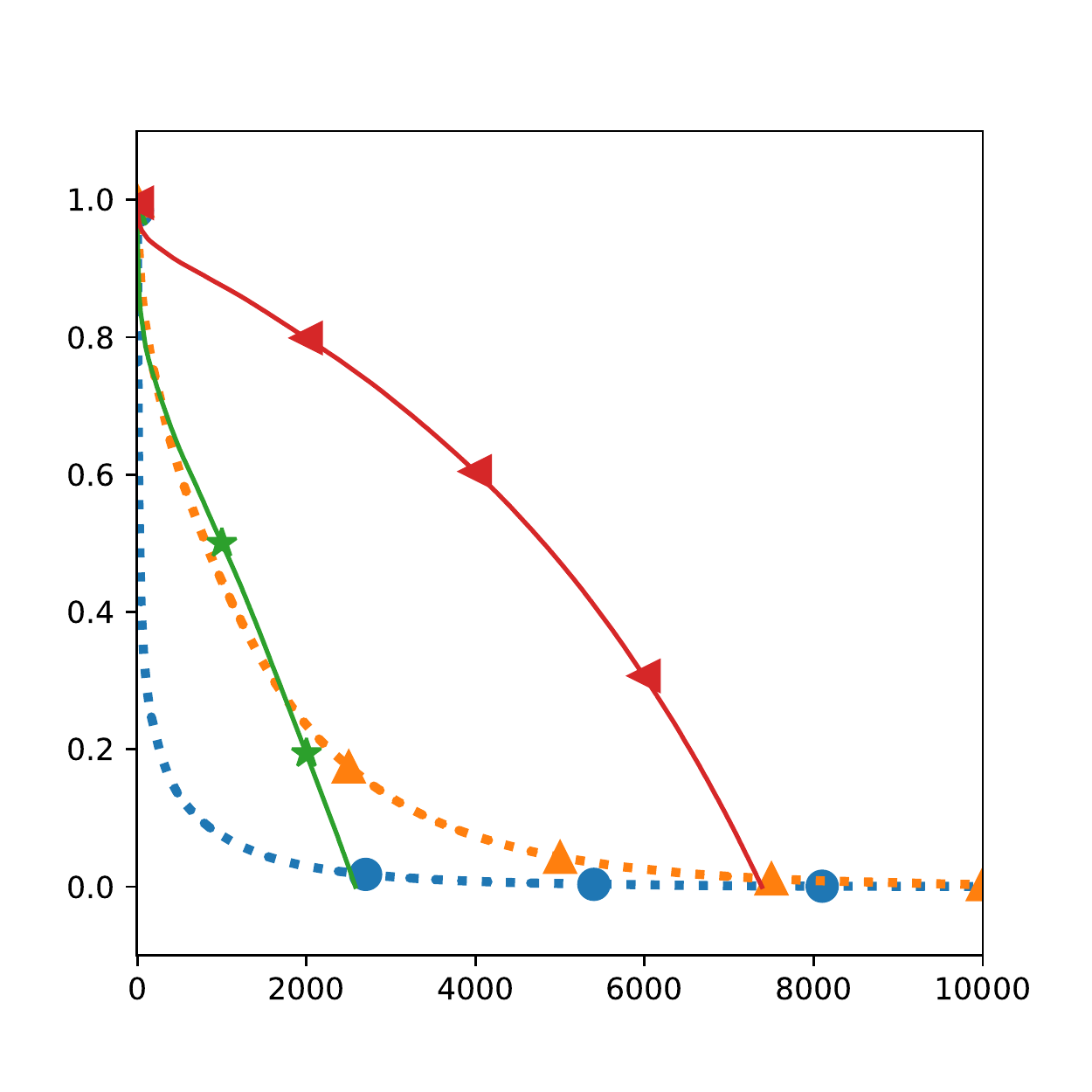} }}%
    \caption{Geometric Transformation for ML-20M.}%
    \label{fig:example}%
\end{figure}

\noindent{\bf The Geometric Transformation}
From the matrix factorization perspective, the Formulas ~\ref{eq:regressionmf} and ~\ref{eq:mf} essentially tells us that these two different approaches both scale-down the singular values of the user-item interaction matrix (binary) with slightly different manner: $\frac{1}{1+\lambda/\sigma_i^2}$ for low-rank regression and $1-\lambda/\sigma_i$ for matrix factorization. Figure~\ref{fig:ml-20m-sigma} illustrates the compressed singular value for {\em ML-20M} dataset with LRR corresponds to the low-rank regression and MF corresponds to the matrix factorization. Figure~\ref{fig:ml-20m-pca} illustrates the compression ratios, which also has a direct geometric explanation: if we consider both approaches as the linear auto-encoder (or regression), as being described in Formulas~\ref{eq:regressionpca} and ~\ref{eq:mfpca}, then the comprehension ratios directly scale down the coordinates ($XV_k$) for the principal component basis in  $V_k^T$. 

% \begin{figure}
%     \centering
%     \includegraphics[width=\linewidth]{Figures/sigma_ml-20m.pdf}
%     \caption{ml-20m, comparing $\sigma$}
%     \label{fig:ml-20m-sigma}
% \end{figure}

\noindent{\bf The Effective Range of $\lambda$ and $\lambda^\prime$ and Latent Dimension $k$}
For the regression based approach, we note that when $\lambda<\sigma_i^2$, then there is relatively small effect to scale down $\sigma_i$ (or any other singular values larger than $\sigma_i$). 
Given this, $\lambda$ tends to be quite big, and it has close to binary effect: let $\sigma_i^2 \geq \lambda_i > \sigma_{i+1}^2$, then any $\sigma_j<\sigma_i$ has small effect, and  for any $\sigma_j>\sigma_i$, then, the reduction will become more significant (proportionally). Typically, $\lambda$ is within the range of the first few singular values (in other words, $i$ is very small, and $\lambda$ is quite large). 

For the matrix factorization, assuming we consider the $k$ dimensional latent factors, then we note that $\sigma_k> \lambda^\prime$, which effectively limit the range of $\lambda^\prime$. 
Furthermore, we notice that since each singular value is reduced by the same amount $\lambda$, which makes the latent dimensions with smaller singular values are even less relevant than their original value. Thus, this leads to the matrix factorization typically has a relatively small number of dimensions to use (typically less than $1000$). 

For the low-rank regression, as the singular value reduced, its proportion $(\frac{1}{1+\lambda/\sigma_i^2})$ will also be reduced down, the same as the matrix factorization. However, unlike the regularized matrix factorization approach whose absolute reduction may vary: 
\begin{equation}
\Delta_i=\sigma_i-\frac{\sigma_i}{1+\lambda/\sigma_i^2} = \frac{\lambda}{\sigma_i+\lambda/\sigma_i}
\end{equation}
Interesting, when $\sigma_i$ is both very large or very small, its absolute reduction are fairly small, but it may reduce those in between more. Thus, this effectively enables the low-rank approaches to use more latent dimensions, thus larger $k$ (typically larger than $1000$). 

Now, an interesting {\em conjecture} of an optimal set of $\lambda_i$ in Equation~\ref{eq:regressionlambda}, is that they should help (relatively) scale-down those large principal components and help (relatively) scale-up those smaller principal components for better performance. However, how can we search the optimal set of $\lambda_i$ for a large number of $k$? We will introduce a learning algorithm for this purpose in Subsection~\ref{learning}, and then utilize that to study the conjecture (through an experimental study in Section~\ref{experiments}). This help provide a better understanding of the adjustment of singular values.  

\noindent{\bf Model Complexity}
For both types of models, we observe that the gram matrix $X^TX$ serves as the sufficient statistics. This indeed suggest that the complexities of both models (number of effective parameters~\cite{hastie@statisticallearning}) will have no higher than $X^T X$. In the past, we typically consider $P$ and $Q$ (together) are defined as the model parameters for matrix factorization. Thus, the common assumption is that MF has model complexities ($O(mk+nk)$). However, the above analysis based on linear autoencoder/regression  perspective, shows that both models essentially only have $V_k$ (together with the scaled principal components). (See Equations~\ref{eq:regressionpca} and ~\ref{eq:mfpca}) for $W$ estimation. Thus, their model complexity are both $O(nk)$ (but with different $k$ for different models). 

Now relating to the aforementioned discussion on the latent dimension factor, we can immediately observe the model complexity of the basic low-rank regression actually have higher complexity than its corresponding matrix factorization model (as the former can allow larger $k$ than the latter). 

Note that for the real-world recommendation models, the matrix factorization will utilize the weight matrix $C$ (equation \ref{eq:WMFALS}) to increase model complexity (which does not have closed-form solution~\cite{hu2008collaborative}). However, due to the alternating least square solution, its number of effective parameters will remain at $O(nk)$. Thus, it is more restricted and less flexible than the regression based approaches.

%% file: text/Boost.tex
\section{Parameter Search and Nearby Models}
\label{learning}

In this section, we first aim to identify a set of {\em optimal} (hyper)parameters $\{\lambda_i:1 \leq i \leq k\}$ for the closed-form solution~\ref{eq:regressionlambda}. Clearly, when the parameter number is small, we can deploy a grid search algorithm as illustrated in Algorithm~\ref{alg:pca} for search $k$ and $\lambda$ for the closed-form in Equation~\ref{eq:regressionpca}. However, for a large number of (hyper)parameters, we have to resort to new approaches (Subsection~\ref{parametersearch}). 
Furthermore, once the parameter searching algorithm is available, we consider to utilize it for searching {\em nearby models} for an existing model (Subsection~\ref{boosting}). As we mentioned before, this can be considered as a post-model fitting exploration in statistical learning~\cite{guan2018post}. 

%the following question: can this method be applied in other models to assistant their parameter search? Since most of the models have their dedicated algorithms for model optimization, we consider to {\em augment} the linear model with a restricted set of parameters, and then search through the augmented parameter space. Why this is useful? This will help us to have better understanding how close the (existing) optimized linear models can deliver for recommendation. 

\begin{algorithm}[t]
\caption{
Hyperparameter Search for Formula~\ref{eq:regressionpca} }\label{alg:pca}
%\begin{small}
%\begin{scriptsize}
\begin{flushleft}
        \textbf{INPUT:} 
        Hyperparameter candidate lists: $\lambda_l$, $k_l$,
        user-item binary matrix $X$.\\
        \textbf{OUTPUT:} Model performance for all hyperparameter $\lambda_l$, $k_l$ combinations.
\end{flushleft}
\begin{algorithmic}[1]
\STATE $X^TX= V \Sigma^T \Sigma V^T $ \text{ (Eigen Decomposition) }
\FORALL{$\lambda \in$ $\lambda$ list} 

    \STATE\parbox[t]{200pt}{$\Delta \coloneqq (\Sigma^T\Sigma + \lambda I)^{-1}\Sigma^T\Sigma = diag(\frac{d^2_1}{d^2_1+\lambda},\dots,\frac{d^2_n}{d^2_n+\lambda})$}
    \FORALL{$k \in$ $k$ list} 
    \STATE ${}$\hspace{2em} $\Delta_k\leftarrow$ first $k$ columns and rows of $\Delta$
    \STATE ${}$\hspace{2em} $V_k\leftarrow$ first $k$ columns of $V$
    \STATE ${}$\hspace{2em} $W_k\leftarrow$   $V_k\Delta_k V^T_k$
    \STATE ${}$\hspace{2em}evaluate($W_k$) based on nDCG and/or Recall@K
    \ENDFOR
\ENDFOR
\end{algorithmic}
%\end{small}
%\end{scriptsize}
\end{algorithm}

\subsection{Parameter Search}
\label{parametersearch}
In this subsection, we assume the closed-form solution in Equation~\ref{eq:regressionpca} ($
\widehat{W} = V_k diag(\frac{\sigma^2_1}{\sigma^2_1+\lambda},\dots,\frac{\sigma^2_n}{\sigma^2_k+\lambda}) V_k^T$) has optimized hyperparameters $\lambda$ and $k$ through the grid search algorithm Algorithm~\ref{alg:pca}. Our question is how to identify optimized parameter $\lambda_1, \cdots \lambda_k$ in Equation~\ref{eq:regressionlambda}: $
\widehat{W} = V_k  diag(\frac{1}{1+\lambda_1/\sigma^2_1},\dots,\frac{1}{1+\lambda_n/\sigma^2_n})  V_k^T$. 

The challenge is that the dimension (rank) $k$ is fairly large and the typical (hyper)parameter search cannot work in such high dimensional space~\cite{randomhyper12}. Also, as we have the closed-form, it does not make sense to utilize the (original) criterion such as Equation~\ref{eq:lowrank} for optimization. Ideally, we would like to  evaluate the accuracy of any parameter setting (such as in Algorithm~\ref{alg:pca}) based on nDCG or AUC~\cite{charubook}. 
Clearly, for this high dimensional continuous space, this is too expensive.  To deal with this problem, we consider to utilize the {\em BPR} loss function which can be considered as a continuous analogy of AUC~\cite{bpr09}, and parameterize $\lambda_i$ with a search space centered around the optimal $\lambda$ discovered by Algorithm~\ref{alg:pca}:
\begin{equation*}
    \lambda_i(\alpha_i)=\lambda+ c \times tanh(\alpha_i), 
\end{equation*}
where $c$ is the search range, which is typically a fraction of $\lambda$ and $\alpha_i$ is the parameter to be tuned in order to find the right $\lambda_i$. Note that this method effectively provides a bounded search space $(\lambda-c, \lambda+c)$ for each $\lambda_i$. 

Given this, the new objective function based on BPR is:  
\begin{equation*}
    \mathcal{L}=\sum_{u,i\in X_u^+,j \in X_u^-} -\log( \delta(t x_u  (W(\alpha_1,\cdots,\alpha_k)_{*i}-W(\alpha_1,\cdots,\alpha_k)_{*j}))  
\end{equation*}
where $W(\alpha_1,\cdots,\alpha_k)=V_k  diag(\frac{\sigma^2_1}{\sigma^2_1+\lambda_1(\alpha_i)},\dots,\frac{\sigma^2_n}{\sigma^2_n+\lambda_n(\alpha_n)})  V_k^T$, 
and $W(\alpha_1,\cdots,\alpha_k)_{*i}$ is the $i$-th column of $W$ matrix, $x_u$ is the $u-th$ row of matrix $X$ and $t$ is a scaling constant. 
Here $t$ and $c$ are hyper-parameters for this learning procedure.

Note that this is a non-linear loss function for a linear model and the entire loss function can be directly implemented as a simple neural network, and ADAM (or other gradient descent) optimization procedure can be utilized for optimization. We can also add other optimization such as dropout and explicitly enforcing the zero diagonal of $W$~\cite{Steck_2019}. 

\subsection{Nearby Linear Models}
\label{boosting}

In this subsection, we consider how to further leverage the new learning procedure for other linear models   to help identify the (hyper)parameters. Inspired by the recent efforts of post-model fitting exploration~\cite{guan2018post},  we consider to augment the existing learned $W$ from any existing models (or adding on top of the aforementioned closed-form solution) with two types of parameters:
\begin{equation}
\label{eq:augment}
\begin{split}
    W_{HT} &= diagM(H) \cdot W \cdot diagM(T) \\
    W_{S} &= S \odot \widehat{W} \odot (\widehat{W}\geq t)
\end{split}
\end{equation}
where $H=(\delta(h_1), \cdots \delta(h_n))$ and $T=(\delta(t_1),\cdots,\delta(t_n))$ are the {\em head} and {\em tail} vectors with values between $0$ and $1$ (implemented through sigmoid function). We also refer to the diagonal matrices $diagM(H)$ and $diagM(T)$ as the head and tail matrices. Basically, these diagonal matrices $diagM(H)$ and $diagM(T)$ help re-scale the row and column vectors in $W$. 
Furthermore, $S=(\delta(s_{ij}))$ is a matrix with values between $0$ and $1$ (implemented through sigmoid function). Finally, $\widehat{W}\geq t$ is a boolean matrix for sparsification: when $\widehat{W}_{ij}>t$, its element is $1$, otherwise, it it zero.  Thus, this augmented model basically consider to sparsify the learned similar matrix $W$ and re-scale its remaining weights. 
Note that both $W_{HT}$ and $W_S$ can be considered as the {\em nearby} models for the existing models with learned $\widehat{W}$. 
Note that studying these models can also help us understand how close these available learner models are with respect to their limit for recommendation tasks. Since the optimization is more close to the ``true'' objective function, it helps us to squeeze out any potential better models near these existing models. 
In Section~\ref{experiments}, we will experimentally validate if there is any space for improvement based on those simple augmented learning models.

%% file: text/eval.tex
\vspace*{-1.0ex}
\section{Experimental Results}
\label{experiments}

% \usepackage{multirow}
% \usepackage{graphicx}
% \usepackage{multirow}
% \usepackage{graphicx}

In this section, we experimentally study the basic linear models as well as the (hyper)parameter search algorithms and its applications to the nearby models. Note that our goal here is not to demonstrate the superiority of these basic/closed-form solutions, but to show they can fare well against the state-of-the-art linear models. This can thus help validate using these basic models to study these advanced linear models~\cite{Steck_2019,DBLP:conf/nips/Steck20,hu2008collaborative}. 
Specifically, we aim to answer: 

%the following questions:

\begin{itemize}[leftmargin=*,noitemsep,nolistsep]
    \item (Question 1) How do the basic regression and matrix factorization based models (and their closed-form solutions) compare against the state-of-the-art linear models? Also we hope to compare the two basic models (using their closed-form solutions) to help provide evidence if the matrix factorization approaches have inherently disadvantage for the implicit recommendation task.  
    
    \item (Question 2) How can the learning algorithm to help search the optimal parameter for the closed-form solution of Equation ~\ref{eq:regressionlambda} as well as its augmented models (adding both head and tail matrices)? How does the (augmented) closed-form solution perform against the state-of-the-art methods? We are also interested in understanding how the learned $\{\lambda_i\}$ parameters look like with respect to the constant $\lambda$. 
    
   \item (Question 3)  How does the nearby models based on the head and tail matrices $W_{HT}$ and sparsification $W_S$ introduced in Subsection~\ref{boosting} perform? Can any existing state-of-the-art linear models  be boosted by searching through the augmented nearby models?  
\end{itemize}

% $(which potentially provides the lower bound for a specific algorithm on a given dataset, based on Beta distribution)

\noindent{\bf Experimental Setup:}
We use  three  commonly used datasets for recommendation studies: MovieLens 20 Million (ML-20M) \cite{ml20m}, Netflix Prize (Netflix) \cite{netflix}, and the Million Song Data (MSD)\cite{msddataset}.  The characteristics of first two datasets are in the bottom of Table~\ref{table: maintable}. The characteristics of the  third dataset and its results is in Appendix. 

For the state-of-the-art recommendation algorithms, we consider the following: ALS~\cite{hu2008collaborative} for matrix factorization approaches, SLIM~\cite{slim01}, EASE~\cite{Steck_2019}, and EDLAE~\cite{DBLP:conf/nips/Steck20} for regression models, CDAE ~\cite{cdae16} and MultiVAE~\cite{liang2018variational} for deep learning models.  For most of the experiment settings, we follow ~\cite{liang2018variational,Steck_2019,DBLP:conf/nips/Steck20} for the {\em strong generalization} by splitting the users into training, validation and tests group. 
Also following ~\cite{liang2018variational,Steck_2019,DBLP:conf/nips/Steck20}, we report the results using metrics $Recall@20$, $Recall@50$ and $nDCG@100$. 

Finally, note that our code are openly available (see Appendix).

% \usepackage{graphicx}
\begin{table}[]
\resizebox{\linewidth}{!}{%
\begin{tabular}{|c|c|c|c|c|}
\hline
ML-20M &
  \begin{tabular}[c]{@{}c@{}}EASE\\ $\lambda$=400\end{tabular} &
  \begin{tabular}[c]{@{}c@{}}LRR\\ k =2K, $\lambda$ = 10K\end{tabular} &
  \begin{tabular}[c]{@{}c@{}}MF\\ k = 1K, $\lambda$ = 50\end{tabular} &
  \begin{tabular}[c]{@{}c@{}}WMF(ALS)\\ k = 100, C = 10, $\lambda$ = 1e2\end{tabular} \\ \hline
Recall@20 & \textbf{0.39111}  & 0.37635           & 0.36358 & 0.36327  \\ \hline
Recall@50 & \textbf{0.52083} & {0.51144} & 0.50069  & 0.50232\\ \hline
nDCG@100  & \textbf{0.42006} & 0.40760          & 0.39187 & 0.39314 \\ \hline
\end{tabular}%
}
\caption{ML-20M: Basic Model Evaluation}
\label{tab:ml-20mbasic}
\vspace*{-6.0ex}
\end{table}

\begin{table}[]
\begin{small}
\resizebox{\linewidth}{!}{%
\begin{tabular}{|c|c|c|c|c|}
\hline
Netflix &
  \begin{tabular}[c]{@{}c@{}}EASE\\ $\lambda$= 1000\end{tabular} &
  \begin{tabular}[c]{@{}c@{}}LRR\\ k =3K, $\lambda$ = 40K\end{tabular} &
  \begin{tabular}[c]{@{}c@{}}MF\\ k=1K, $\lambda$=100\end{tabular} &
  \begin{tabular}[c]{@{}c@{}}WMF(ALS)\\ k=100, C = 5, $\lambda$=1e2\end{tabular} \\ \hline
Recall@20 & \textbf{0.36064}  & 0.3478 & 0.33117  & 0.3213 \\ \hline
Recall@50 & \textbf{0.44419}  & 0.4314   & 0.41719   & 0.40629 \\ \hline
nDCG@100  & \textbf{0.39225} & 0.38018 & 0.36462 & 0.35548  \\ \hline
\end{tabular}%
}
\end{small}
\caption{Netflix: Basic Model Evaluation}
\label{tab:netflixbasic}
\vspace{-6.0ex}
\end{table}

\noindent{\bf Basic Model Evaluation:} In this experiment, we aim to evaluate the the closed-form (Formulas~\ref{eq:regressionpca}, referred to as $LRR$ and ~\ref{eq:mfpca}, referred to as $MF$)  of the two basic models (Equations~\ref{eq:lowrank} and  ~\ref{eq:matrixfactorization}). We compare them against the state-of-the-art regression model EASE~\cite{Steck_2019} and ALS~\cite{hu2008collaborative}. Since this is mainly for evaluating their prediction capacity (not on how they perform on the real world environment), here we utilize the leave-one-out method to evaluate these models. Note that this actually provides an advantage to the matrix factorization approaches as they prefer to learn the embeddings (latent factors) before its prediction. 

Tables~\ref{tab:ml-20mbasic} and ~\ref{tab:netflixbasic} show the results for these four linear models on the ml-20m and netflix datasets, respectively. We perform a grid search for each of these models (the grid search results are reported in Appendix), and report their better settings (and results) in these tables. From these results, we observe:
(1) Both basic models $LRR$ and $MF$ have very comparable performances against their advanced version. Note that $LRR$ does not have the zero diagonal constraint and use reduced rank regression compared with EASE; and $MF$ does not have the weighted matrix in ALS~\cite{hu2008collaborative}.  This helps confirm the base models can indeed capture the essence of the advanced models and thus our theoretical analysis on these models can help (partially) reflect the behaviors from advanced models.  
(2) Both regression models are consistently and significantly better than the matrix factorization based approaches. This helps further consolidate the observations from other studies~\cite{DacremaBJ21} that the regression methods have the  advantage over the matrix factorization methods. 

\begin{figure}%
\vspace*{-3.0ex}
    \centering
    \subfloat[\centering optimal adjusted singular value ]{{\includegraphics[width=.615\linewidth]{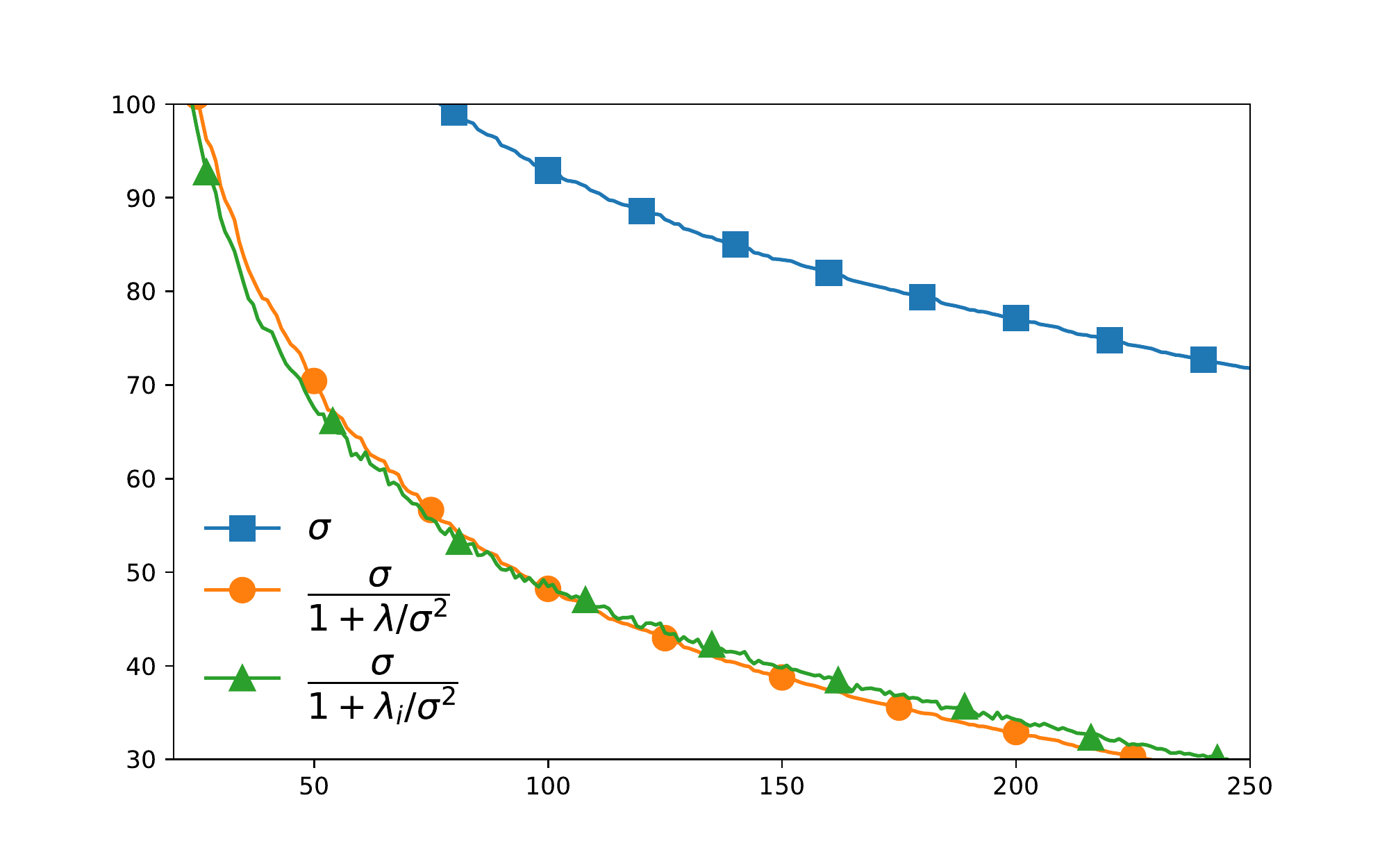} }}%
    \subfloat[\centering optimized (hyper)parameters $\lambda_i$ ]{{\includegraphics[width=.385\linewidth]{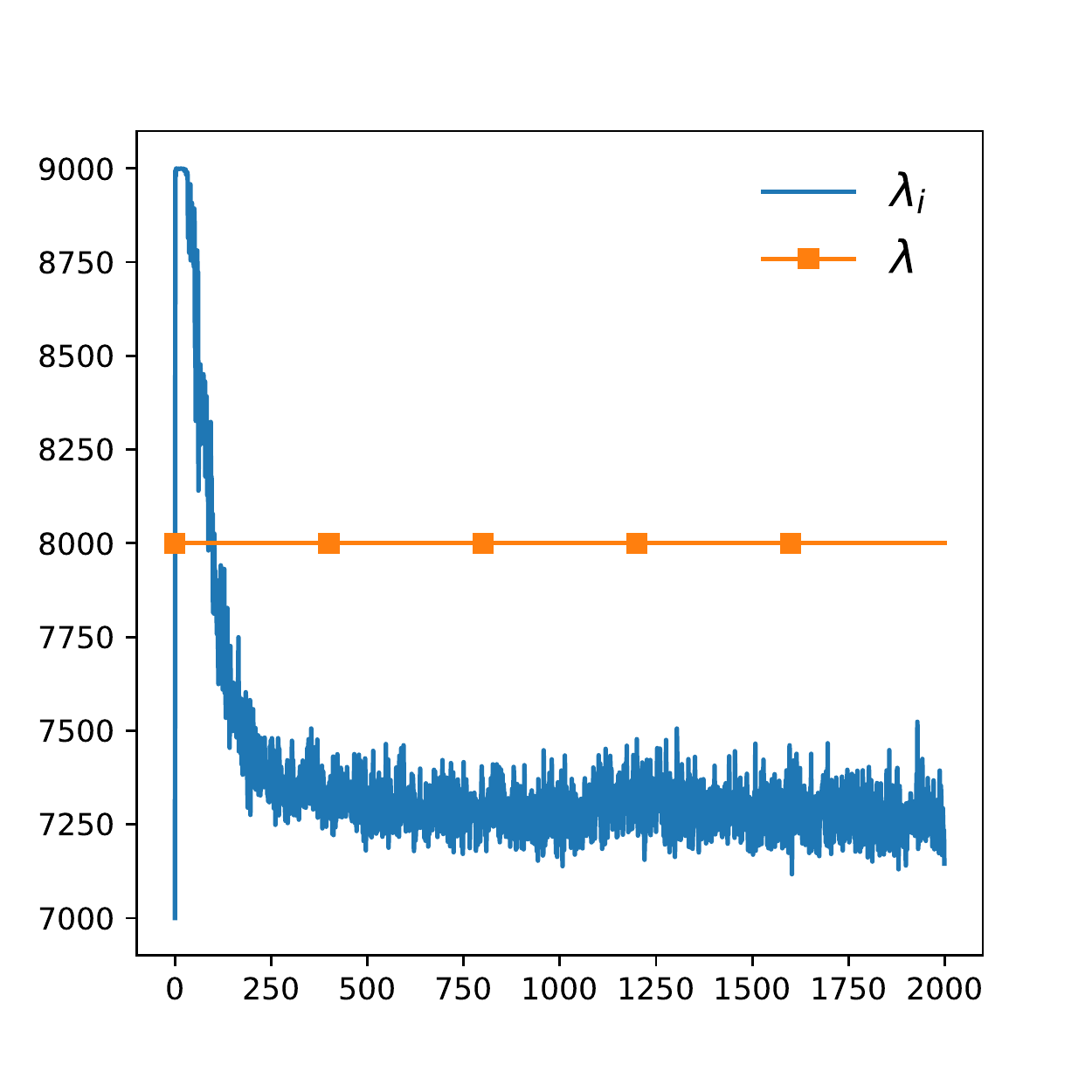} }}%
    \caption{Optimization Results for Formula~\ref{eq:regressionlambda}}%
    \label{fig:parameterresults}%
    \vspace{-5.0ex}
\end{figure}

\begin{table*}[th]
\resizebox{\textwidth}{!}{%
\begin{scriptsize}
\begin{tabular}{|c|c|c|c|c|c|c|c|}
\hline
\multicolumn{2}{|c|}{\multirow{2}{*}{Model}}   & \multicolumn{3}{c|}{ML-20M} & \multicolumn{3}{c|}{Netflix} \\ \cline{3-8} 
\multicolumn{2}{|c|}{}                         & Recall@20 & Recall@50 & nDCG@100 & Recall@20 & Recall@50 & nDCG@100 \\ \hline
\multicolumn{2}{|c|}{LRR (closed-form) } & 0.376     & 0.513     & 0.406    & 0.347     & 0.432     & 0.380    \\ \cline{3-8} 
\multicolumn{2}{|c|}{LRR + $\lambda_i$}             & 0.380     & 0.515     & 0.410    & 0.348     & 0.433     & 0.381    \\ \cline{3-8} 
\multicolumn{2}{|c|}{LRR + $\lambda_i$ + HT}              & 0.386     & 0.520     & 0.418    & 0.351     & 0.435     & 0.384    \\ \cline{3-8} 
\multicolumn{2}{|c|}{LRR + $\lambda_i$ + HT + RMD}          & 0.386     & 0.522     & 0.418    & 0.351     & 0.435     & 0.384    \\ \hline
\multicolumn{2}{|c|}{EDLAE}          & 0.389     & 0.521     & 0.422    & 0.362     & 0.446     & 0.393    \\ \cline{3-8} 
\multicolumn{2}{|c|}{EDLAE + $HT$}              & 0.394     & 0.527     & 0.424    & 0.361     & 0.446     & 0.393    \\ \cline{3-8} 
\multicolumn{2}{|c|}{EDLAE Full Rank}                & 0.393     & 0.523     & 0.424    & 0.364     & 0.449     & 0.397    \\ \cline{3-8} 
\multicolumn{2}{|c|}{EDLAE Full Rank + $HT$}          & 0.395     & 0.527     & 0.426    & 0.364     & 0.449     & 0.396    \\ \cline{3-8} 
\multicolumn{2}{|c|}{EDLAE Full Rank + Sparsification} & 0.394     & 0.526     & 0.423    & 0.365     & 0.450     & 0.397    \\ \hline
\multicolumn{2}{|c|}{SLIM}                     & 0.370     & 0.495     & 0.401    & 0.347     & 0.428     & 0.379    \\ \cline{3-8} 
\multicolumn{2}{|c|}{ALS/WMF}                      & 0.363    & 0.502     & 0.393    & 0.321     & 0.406     & 0.355    \\ \cline{3-8} 
\multicolumn{2}{|c|}{EASE}                     & 0.391     & 0.521     & 0.420    & 0.360     & 0.444     & 0.392    \\ \cline{3-8} 
\multicolumn{2}{|c|}{CDAE}                     & 0.391     & 0.523    & 0.418   & 0.343     & 0.428    & 0.376    \\ \cline{3-8} 
\multicolumn{2}{|c|}{MULT-DAE}                     & 0.387     & 0.524     & 0.419   & 0.344     & 0.438     & 0.380   \\ \cline{3-8} 
\multicolumn{2}{|c|}{MULT-VAE}                 & 0.395     & 0.537     & 0.426    & 0.351     & 0.444     & 0.386    \\ \hline
\multirow{3}{*}{dataset statistics} & \# items & \multicolumn{3}{c|}{20108}  & \multicolumn{3}{c|}{17769}   \\
               & \# users                      & \multicolumn{3}{c|}{136677}      & \multicolumn{3}{c|}{463435}      \\
               & \# interactions               & \multicolumn{3}{c|}{10 millions}       & \multicolumn{3}{c|}{57 millions}       \\ \hline
\end{tabular}%
\end{scriptsize}
}
\caption{The performance comparison between different models. $\lambda_i$:learned (hyper)parameters; HT: augmented models with head and tail parameter matrix; RMD: with removing the diagonal matrix (enforcing zero diagonal). For more details of the experimental set-up and model, please refer to the appendix.}
\label{table: maintable}
\end{table*}

\noindent{\bf Optimizing Closed-Form Solutions:}
In this and next experiment, we will follow the strong generalization setting by splitting the users into training, validation and testing groups. The top section of Table~\ref{table: maintable} shows the experimental results of using the closed-form solution (Formula~\ref{eq:regressionlambda}). Here, (1) $LRR(closed-form)$ is the starting point for $\lambda$ being constant; (2) $LRR+\lambda_i$ utilizes the BPR learning algorithm in Subsection~\ref{learning} to search the hyperparameter space; (3) $LRR+\lambda_i+HT$ uses $diagM(H)\cdot W \cdot diagM(T)$ (as the targeted similarity matrix), where $W$ is defined in Formula~\ref{eq:regressionlambda} (here the optimization will simultaneously search hyperparameters $\{\lambda_i\}$ and head ($H$), tail ($T$) vectors; (4) finally, $LRR+\lambda_i+HT+RMD$ further enforces the zero diagonal constraints. We also add dropout (with dropout rate $0.5$) for the model training for models ($2-4$). 

We observe the variants of the closed-form solutions are comparable against the state-of-the-art linear models and deep learning models. For instance, on ML-20M, $LRR+\lambda_i+HT+RMD$ reports $0.522$ $Recall@50$, is among the best for the existing linear models (without additional boosting from the augmented nearby models). 

Finally, Figure~\ref{fig:parameterresults} illustrates the parameter search results for  Formula~\ref{eq:regressionlambda} from the learning algorithm.  Specifically, Figure~\ref{fig:parameterresults} (a) shows how the singular value are adjusted vs the compressed singular value for a constant $\lambda=8000$ (Formula~\ref{eq:regressionpca}). We provide $c = 1000$ to allow each individual $\lambda_i$ search between $7000$ to $9000$. Figure~\ref{fig:parameterresults} (b) shows the search results for the parameters $\lambda_i$. 
As we conjectured, we can see that the initial $\lambda_i$ is quite large which leads to smaller singular values compared with adjusted singular value from Formula~\ref{eq:regressionpca}. Then the parameters $\lambda_i$ reduces which make the smaller singular values reduced less. This can help more (smaller) singular values to have better presence in the final prediction.  

\noindent{\bf Nearby Models}
In this experiment, we augment the latest regression models  EDLAE (full rank and reduced rank) ~\cite{DBLP:conf/nips/Steck20} with additional parameters and apply the parameter learning algorithm to optimize the parameters: 
(1) $EDLAE$ is the original reduced rank regression with rank $k=1000$; (2) $EDLAE+HT$ corresponds to the augmented model with head and tail matrices, $W_{HT}$ from Formula~\ref{eq:augment}; (3) $EDLAE\ Full\ Rank$ is the original full rank regression; (4) $EDLAE\ Full\ Rank+HT$ applies the head and tail matrices on the learned similarity matrix from $EDLAE\ Full\ Rank$; (5) $EDLAE\ Full\ Rank+Sparsification$ applies the $W_S$ from Formula~\ref{eq:augment}, which sparsifies the similarity matrix of $EDLAE\ Full\ Rank$ with additional parameters in matrix $S$ to further adjust those remaining entries in the similarity matrix. 

The experimental results on ML-20M and Netflix of these augmented (nearby) models are listed in the middle section in Table~\ref{table: maintable}. We can see that on the ML-20M dataset, the $Recall@50$ has close to $1\%$ boost while other metrics has small improvement. This indeed demonstrates the nearby models may provide non-trivial improvement over the existing models. On the Netflix dataset, the nearby models only have minor changes and indicates the originally learned model may already achieve the local optimum. 
 

%% file: text/conclusion.tex
\vspace*{-2.0ex}
\section{Conclusion and Discussion}

In this work, we provide a thorough investigation into the relationship between arguably two of the most important recommendation approaches: the neighborhood regression approach and the matrix factorization approach. We show how they inherently connect with each other as well as how they differ from one another. However, our study mainly focuses on the implicit setting: here the goal is not to recover the original ratings (like in the explicit setting), but to recover a ''likelihood'' (or a preference) of the interaction. Thus, the absolute value/rating is not of interests. In fact, for most of the linear regression models, the predicted value can be very small (more close to zero than one). What matters here is the relative rank of the predicted scores. Thus it helps to use more latent factors to express the richness of user-item interactions. This can be rather different from the rating recovery, which requires the original singular values to be preserved. 
Especially, the current approaches of explicit matrix factorization which often consider only the positive values and thus the methodology developed in this work cannot be immediately applied in this setting. Indeed, Koren and Bell in ~\cite{advancedCF} has analyzed the relationship between neighborhood and factorization models under explicit settings. It remains to be seen whether the insights gained here can be applied to the explicit setting. 

%Along the way, we resolve the questions on how their (hyper)parameters should be chosen and help clarify a somewhat counter-intuitive observation on the model complexity: though on the surface the matrix factorization approaches have more parameters $O((m+n)k)$ where as the regression approaches are no more than $O(n^2)$, it turns out the linear regressions tend to have higher model complexities (even for a small $n$). 

Also, we would like to point out that this is the first work to investigate the nearby linear models. We consider two basic models which utilize limited additional parameters to help explore the additional models. An interesting question is whether we can explore more nearby models. 

%and if there is any way to provide an upper bound of accuracy for the linear (or non-linear) recommendation models on a given datasets. 

Finally, we note that the theoretical models need eigen decomposition which makes them infeasible for the real-world datasets with millions of items. But our purpose here is to leverage such models to help understand the tradeoffs and limitations of linear models, not to replace them.  We hope what being revealed in this work can help design better linear and nonlinear models for recommendation.

%% file: text/appendix_new.tex
\appendix
\label{appendix}
\section{Reproducibility}
Generally, We follow the (strong generalization) experiment set-up in ~\cite{liang2018variational,DBLP:conf/nips/Steck20} and also the pre-processing of the three public available datasets, MovieLens 20 Million (ML-20M) \cite{ml20m}, Netflix Prize (Netflix) \cite{netflix}, and the Million Song Data (MSD)\cite{msddataset}.

\subsection{Experimental Set-up for Table \ref{tab:ml-20mbasic} and \ref{tab:netflixbasic}}
For the experiment of table \ref{tab:ml-20mbasic} and \ref{tab:netflixbasic}, we utilize the strong generalization protocol for EASE \cite{Steck_2019} and LRR methods. For Matrix Factorization based methods (MF and WMF/ALS), they are trained for with data (except the items to be evaluated in validation and test sets)  Note that this actually provides an advantage to the matrix factorization approaches as they prefer to learn the embeddings (latent factors) before its prediction. The experiment results present in table \ref{tab:ml-20mbasic} and \ref{tab:netflixbasic} are obtained by parameter grid search over the validation set according to $nDCG@100$, the same as \cite{liang2018variational}. The searching results are listed as following : table \ref{tab:ml20m:mf:search}, table \ref{tab:netflix:mf:search},
table \ref{tab:ml20m:lrr:search} and table \ref{tab:netflix-lrr-search}.

% \usepackage{multirow}
% \usepackage{graphicx}
\begin{table}[H]
\resizebox{\linewidth}{!}{%
\begin{tabular}{|c|c|c|c|c|c|c|c|c|}
\hline
\multicolumn{2}{|c|}{\multirow{2}{*}{}} & \multicolumn{7}{c|}{$\lambda$}                                                \\ \cline{3-9} 
\multicolumn{2}{|c|}{}                  & 0       & 10      & \textbf{50}      & 100     & 200     & 500    & 1000   \\ \hline
\multirow{5}{*}{k}    & 128             & 0.29874 & 0.30951 & 0.36488          & 0.37826 & 0.30121 & 0.1901 & 0.1901 \\ \cline{2-9} 
                      & 256             & 0.22911 & 0.25104 & 0.37504          & 0.37826 & 0.30121 & 0.1901 & 0.1901 \\ \cline{2-9} 
                      & 512             & 0.1546  & 0.19782 & 0.39682          & 0.37826 & 0.30121 & 0.1901 & 0.1901 \\ \cline{2-9} 
                      & \textbf{1000}   & 0.09177 & 0.18242 & \textbf{0.39893} & 0.37826 & 0.30121 & 0.1901 & 0.1901 \\ \cline{2-9} 
                      & 1500            & 0.06089 & 0.20776 & 0.39893          & 0.37826 & 0.30121 & 0.1901 & 0.1901 \\ \hline
\end{tabular}%
}
\caption{ML-20M, MF, parameter search}
\label{tab:ml20m:mf:search}
\end{table}

\begin{table}[H]
\resizebox{\linewidth}{!}{%
\begin{tabular}{|c|c|c|c|c|c|c|c|c|}
\hline
\multicolumn{2}{|c|}{\multirow{2}{*}{}} & \multicolumn{7}{c|}{$\lambda$}                                                  \\ \cline{3-9} 
\multicolumn{2}{|c|}{}                  & 0       & 10      & 50      & \textbf{100}     & 200     & 500     & 1000    \\ \hline
\multirow{5}{*}{k}    & 128             & 0.31297 & 0.31542 & 0.32607 & 0.34103          & 0.34132 & 0.25831 & 0.17414 \\ \cline{2-9} 
                      & 256             & 0.25036 & 0.25536 & 0.28659 & 0.33521          & 0.34132 & 0.25831 & 0.17414 \\ \cline{2-9} 
                      & 512             & 0.17485 & 0.18314 & 0.26081 & 0.35157          & 0.34132 & 0.25831 & 0.17414 \\ \cline{2-9} 
                      & \textbf{1000}   & 0.12036 & 0.13766 & 0.28868 & \textbf{0.36414} & 0.34132 & 0.25831 & 0.17414 \\ \cline{2-9} 
                      & 1500            & 0.09147 & 0.12449 & 0.32103 & 0.36414          & 0.34132 & 0.25831 & 0.17414 \\ \hline
\end{tabular}%
}
\caption{Netflix, MF, parameter search}
\label{tab:netflix:mf:search}
\end{table}

\begin{table}[H]
\resizebox{\linewidth}{!}{%
\begin{tabular}{|c|c|r|r|r|r|r|r|r|}
\hline
\multicolumn{2}{|c|}{\multirow{2}{*}{}} & \multicolumn{7}{c|}{$\lambda$}                                                  \\ \cline{3-9} 
\multicolumn{2}{|c|}{} &
  \multicolumn{1}{c|}{8000} &
  \multicolumn{1}{c|}{9000} &
  \multicolumn{1}{c|}{\textbf{10000}} &
  \multicolumn{1}{c|}{11000} &
  \multicolumn{1}{c|}{12000} &
  \multicolumn{1}{c|}{13000} &
  \multicolumn{1}{c|}{14000} \\ \hline
\multirow{3}{*}{k}    & 1000            & 0.41063 & 0.41273 & 0.41432          & 0.41476 & 0.41515 & 0.41513 & 0.41478 \\ \cline{2-9} 
                      & \textbf{2000}   & 0.41332 & 0.41469 & \textbf{0.41533} & 0.41509 & 0.41499 & 0.41455 & 0.41394 \\ \cline{2-9} 
                      & 3000            & 0.41282 & 0.41397 & 0.41473          & 0.4146  & 0.41452 & 0.41413 & 0.41347 \\ \hline
\end{tabular}%
}
\caption{ML-20M, LRR, parameter search}
\label{tab:ml20m:lrr:search}
\end{table}

% \usepackage{multirow}
% \usepackage{graphicx}
\begin{table}[H]
\resizebox{\linewidth}{!}{%
\begin{tabular}{|c|c|r|r|r|r|r|r|}
\hline
\multicolumn{2}{|c|}{\multirow{2}{*}{}} & \multicolumn{6}{c|}{$\lambda$}                                        \\ \cline{3-8} 
\multicolumn{2}{|c|}{} &
  \multicolumn{1}{c|}{10000} &
  \multicolumn{1}{c|}{20000} &
  \multicolumn{1}{c|}{30000} &
  \multicolumn{1}{c|}{\textbf{40000}} &
  \multicolumn{1}{c|}{50000} &
  \multicolumn{1}{c|}{60000} \\ \hline
\multirow{3}{*}{k}    & 2000            & 0.33632 & 0.37139 & 0.37856 & 0.37942          & 0.37828 & 0.37644 \\ \cline{2-8} 
                      & \textbf{3000}   & 0.34905 & 0.37441 & 0.37934 & \textbf{0.37949} & 0.37807 & 0.37617 \\ \cline{2-8} 
                      & 4000            & 0.35184 & 0.37468 & 0.37919 & 0.37931          & 0.37786 & 0.37601 \\ \hline
\end{tabular}%
}
\caption{Netflix, LRR, parameter search}
\label{tab:netflix-lrr-search}
\end{table}

\subsection{Experimental Set-up for Table \ref{table: maintable}}

In table \ref{table: maintable}, for LRR (closed-form) model ( described in equation \ref{eq:regressionpca}). For ML-20M dataset, we set $k=2000$, $\lambda = 8000$, $c = 1000$ (used to control range of weighted $\lambda_i$). For Netflix dataset the $\lambda = 8000$ , $\lambda = 40000$, $c = 5000$ . Noting that these hyper-parameters are not set as optimal ones (described in table \ref{tab:ml20m:lrr:search}, table \ref{tab:netflix-lrr-search}), which won't affect our claims. For EDLAE (including full rank) model, we obtain the similarity matrix by running the code from \cite{DBLP:conf/nips/Steck20}. For WMF/ALS model and EASE model, we set the hyper-parameters as table \ref{tab:ml-20mbasic} and table \ref{tab:netflixbasic}. Other models' results are obtained form \cite{liang2018variational}, \cite{Steck_2019} and \cite{DBLP:conf/nips/Steck20}.

For fast training augmented model, we sample part of training data. Generally, it takes 2.5 minutes per 100 batch (batch size is 2048) for training.

\subsection{MSD Dataset Results}

The table \ref{tab:msd_data} shows our experiment results carried out on the MSD dataset. Baseline models' results are obtained form \cite{liang2018variational}, \cite{Steck_2019} and \cite{DBLP:conf/nips/Steck20}.

% \usepackage{multirow}
% \usepackage{graphicx}
\begin{table}[H]
\resizebox{\linewidth}{!}{%
\begin{tabular}{|c|c|c|c|c|}
\hline
\multicolumn{2}{|l|}{\multirow{2}{*}{}}               & \multicolumn{3}{c|}{MSD}                          \\ \cline{3-5} 
\multicolumn{2}{|l|}{}                                & Recall@20       & Recall@50       & nDCG@100      \\ \hline
\multicolumn{2}{|c|}{LRR}                             & 0.24769         & 0.33509         & 0.30127       \\ \cline{3-5} 
\multicolumn{2}{|c|}{LRR + $\lambda_i$}                    & 0.25083         & 0.33902         & 0.30372       \\ \hline
\multicolumn{2}{|c|}{EDLAE}                           & 0.26391         & 0.35465         & 0.31951       \\ \cline{3-5} 
\multicolumn{2}{|c|}{EDLAE Full Rank}                 & 0.33408         & 0.42948         & 0.39151       \\ \cline{3-5} 
\multicolumn{2}{|c|}{EDLAE Full Rank+HT}              & 0.33423         & 0.43134         & 0.38851       \\ \hline
\multicolumn{2}{|c|}{SLIM}                            & \multicolumn{3}{c|}{did not finished in \cite{slim01}} \\ \cline{3-5} 
\multicolumn{2}{|c|}{WMF}                             & 0.211           & 0.312           & 0.257         \\ \cline{3-5} 
\multicolumn{2}{|c|}{EASE}                            & 0.333           & 0.428           & 0.389         \\ \cline{3-5} 
\multicolumn{2}{|c|}{CDAE}                            & 0.188           & 0.283           & 0.237         \\ \cline{3-5} 
\multicolumn{2}{|c|}{MULT-DAE}                        & 0.266           & 0.363           & 0.313         \\ \cline{3-5} 
\multicolumn{2}{|c|}{MULT-VAE}                        & 0.266           & 0.364           & 0.316         \\ \hline
\multirow{3}{*}{dataset statistics} & \# items        & \multicolumn{3}{c|}{41140}                        \\ \cline{2-5} 
                                    & \# users        & \multicolumn{3}{c|}{571355}                       \\ \cline{2-5} 
                                    & \# interactions & \multicolumn{3}{c|}{34 millions}                  \\ \hline
\end{tabular}%
}
\caption{The performance comparison between models on MSD dataset.}
\label{tab:msd_data}
\end{table}

% \usepackage{multirow}
% \usepackage{graphicx}
% \begin{table}[H]
% \resizebox{\linewidth}{!}{%
% \begin{tabular}{|c|c|c|c|c|}
% \hline
% \multicolumn{2}{|l|}{\multirow{2}{*}{}}               & \multicolumn{3}{c|}{MSD}         \\ \cline{3-5} 
% \multicolumn{2}{|l|}{}                                &nDCG@100  & Recall@20 &  Recall@20\\ \hline
% \multicolumn{2}{|c|}{EDLAE}                           & 0.31951   & 0.26391   & 0.35465  \\ \hline
% \multicolumn{2}{|c|}{EDLAE Full Rank}                 & 0.39151   & 0.33408   & 0.42948  \\ \hline
% \multicolumn{2}{|c|}{EDLAE Full Rank+HT}              & 0.38851   & 0.33423   & 0.43134  \\ \hline
% \multirow{3}{*}{dataset statistics} & \# items        & \multicolumn{3}{c|}{41140}       \\ \cline{2-5} 
%                                     & \# users        & \multicolumn{3}{c|}{571355}      \\ \cline{2-5} 
%                                     & \# interactions & \multicolumn{3}{c|}{34mil}       \\ \hline
% \end{tabular}%
% }
% \caption{MSD dataset}
% \label{tab:msd_data}
% \end{table}

% \subsection{Anonymous Code}
% \url{https://anonymous.4open.science/repository/297cbd26-4197-459c-bca5-5d4546f3d2f0/}